# Agile Elicitation of Scalability Requirements for Open Systems: A Case Study

*Gunnar Brataas*, SINTEF Digital, Trondheim, Norway
*Antonio Martini*, University of Oslo, Norway
*Geir Kjetil Hanssen*, SINTEF Digital, Trondheim, Norway
*Georg Ræder,* TietoEVRY, Fornebu, Norway

Corresponding author: Gunnar Brataas, SINTEF Digital, P.O. Box 4760 Torgarden, NO-7465 Trondheim, Norway, Gunnar.Brataas@sintef.no

## Abstract

Eliciting scalability requirements during agile software development is complicated and poorly described in previous research. This article presents a lightweight artifact for eliciting scalability requirements during agile software development: the ScrumScale model. The ScrumScale model is a simple spreadsheet. The scalability concepts underlying the ScrumScale model are clarified in this design science research, which also utilizes coordination theory. This paper describes the open banking case study, where a legacy banking system becomes open. This challenges the scalability of this legacy system. The first step in understanding this challenge is to elicit the new scalability requirements. In the open banking case study, key stakeholders from TietoEVRY spent 55 hours eliciting TietoEVRY's open banking project's scalability requirements. According to TietoEVRY, the ScrumScale model provided a systematic way of producing scalability requirements. For TietoEVRY, the scalability concepts behind the ScrumScale model also offered significant advantages in dialogues with other stakeholders.

## Keywords

Performance requirements, software performance engineering, agile software development, non-functional requirements, open systems, design science.[1]

## 1   Introduction

Eliciting scalability requirements during software development is complicated. Conventional software performance engineering takes time and requires deep competence (Becker et al., 2017), both of which are often not available. Time to elicit scalability requirements is scarce, mainly because the industry embraces agile methods intended to be lightweight. Moreover, the agile literature offers little advice on eliciting scalability requirements (Medeiros et al., 2020). In our previous work, we introduced a set of scalability concepts (Brataas and Herbst et al., 2017) and the first agile method for scalability engineering, the ScrumScale method (Brataas et al., 2020). In this article, together with our industrial partner TietoEVRY, we have built and evaluated a lightweight artifact termed the ScrumScale model to estimate scalability requirements. We refer to the *artifact* as a broad term comprising constructs, models, methods, and implementations (Hevner and Chatterjee, 2010), rather than the more specific *artifacts* defined in Scrum, like user stories, post-it notes. These three artifacts — the method, the conceptual framework, and the model — make up the ScrumScale approach to agile scalability engineering, as shown in Figure 1:

- The ScrumScale method describes steps in agile scalability engineering (Brataas et al., 2020).
- The ScrumScale framework contains key scalability concepts (Brataas and Herbst et al., 2017).
- The ScrumScale model rapidly gives an overview of the projected workload requirements. These scalability requirements are expressed as color-coded output parameters, where red means a large scalability risk, yellow a moderate scalability risk, and green a low scalability risk. In Figure 1, we have two scenarios for scalability requirements: *possible* and *extreme*. These two scenarios reflect increasing levels of market optimism and, consequently, workload. For both these two scenarios, we have input parameters. Based on expert evaluation of the resulting scalability requirement for the *possible* scenario, the yellow color represents a moderate scalability risk. In contrast, the scalability requirement for the *extreme* scenario constitutes a high scalability risk, as indicated by a red color. The numbers used in this initial example are random.

---

[1] Abbreviations used in this paper:
- ASD: agile software development
- BACS: Building Application Case Study
- NFR: non-functional requirement
- PSD2: Revised Directive on Payment Services
- SPE: software performance engineering
- TPPs: third-party providers

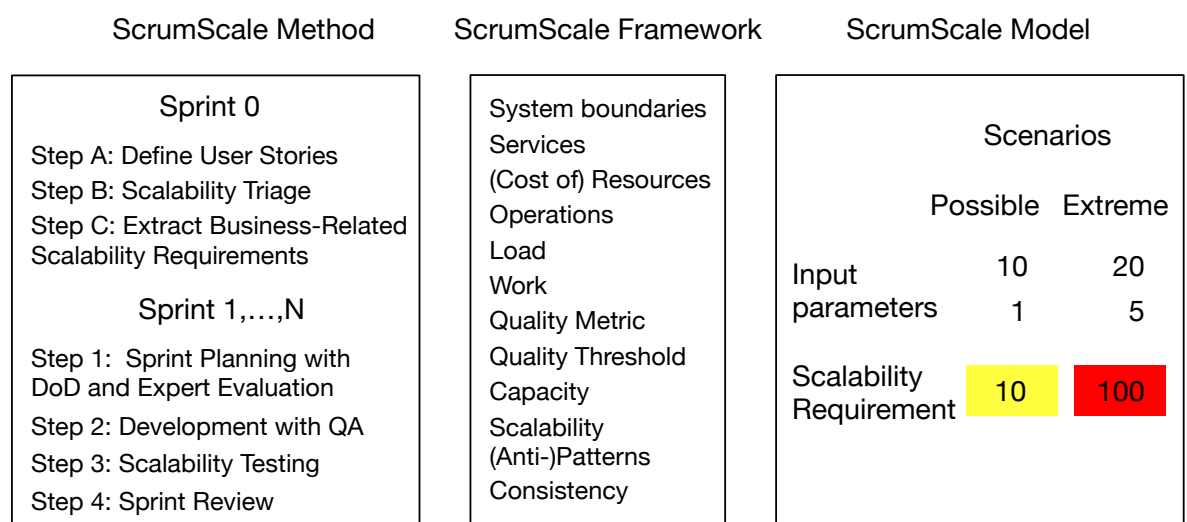

Figure 1: The three ScrumScale artifacts: The method, the framework, and the model.

The ScrumScale model is used to elicit an open system's scalability requirements in this paper. By *open system*, we here mean that the system offers interfaces through which external parties may interact. The ongoing digitalization in many domains may radically affect the workload of stable, well-proven legacy systems. Traditionally closed systems become partially open systems. The workload that has gradually increased for decades and which has been controllable may suddenly boom. Hence, scalability, defined as a system's ability to increase its capacity by consuming more resources (Brataas and Herbst et al., 2017), becomes business-critical. This may incur scalability debt, defined as "architectural technical debt originating by harder scalability requirements because of substantial changes in the business (environment)" (Hanssen et al., 2019). The first step in the timely mitigation of scalability debt is to elicit and analyze scalability requirements. This is especially important in open systems, with potentially abrupt changes in scalability requirements where consequences can be hard to foresee.

In this paper, we present such a case, in which the partly legacy banking system of TietoEVRY[2] is challenged via the mandatory Revised Directive on Payment Services (PSD2). PSD2 requires access to

---

core financial services for third-party providers (TPPs), including competing banks. *Open banking* is a generalization of PSD2, with open access to many financial service APIs. Both PSD2 and open banking may radically increase workloads generated from new third-party solutions. The uncertainty in workload estimation is intensified because of last-minute changes in PSD2 regulations. Because the primary aim of PSD2 is increased competition in financial services, the competition also becomes tougher. As a result, the deployment of new or improved services will happen more frequently. Financial services that do not handle this increased workload, more uncertainty, and intensified competition may be out of business.

As a result, TietoEVRY's capacity for scalability engineering becomes business-critical. In the invitation to the kick-off meeting for this case study in May 2018, the representative for system development processes in TietoEVRY (and the fourth author of this paper) wrote: *«The ScrumScale project has a model for (scalability) requirements analysis (meaning Brataas and Herbst et al., 2017), and it could be useful to analyze in detail what projected load (on the open banking system) would be."* He further states, *"Ideally, we should be able to estimate projected workload per service as an input to performance testing.»* This paper will describe how we helped TietoEVRY elicit scalability requirements — the first critical step toward sound scalability engineering.

When we initiated this case study, we knew that the stakeholders were busy and that new potential scalability risks from new PSD2 regulations were under-focused. Therefore, both the process and the artifact we aimed to establish had to be lightweight. We also knew many different stakeholders had to participate in the scalability requirement elicitation, such as testers, architects, subsystem architects. However, we did not know if it would be feasible to develop a sufficiently lightweight artifact, especially in a cost-effective manner, since our agile scalability engineering method (Brataas et al., 2020) did not specify this. As a result, the first research question was:

**RQ1: Is it feasible to use an iterative and lightweight approach to create an effective artifact to elicit scalability requirements?**

By effective, we mean an artifact that would include crucial knowledge, and at the same time, would be intuitive, easy to learn and use, useful for discussions and sharing across multiple stakeholders. Besides understanding whether it was feasible to develop the artifact using a minimum of time and resources, we wanted to know if they were good enough for the purpose and to find out how successful, or not, our approach was. The second research question therefore was:

**RQ2: How well does our artifact support agile scalability requirement elicitation for open systems?**

This paper has applied a design science process to create a lightweight artifact to elicit scalability requirements. Design science assures that critical business needs are considered and that the developed artifact is continuously monitored and refined and thoroughly evaluated against the initial requirements. We have combined a conceptual scalability framework, the ScrumScale framework, with a process for agile requirement elicitation (based on the ScrumScale method), agile architecture, and coordination theory. Coordination theory becomes useful when creating an artifact that needs to coordinate the key input from various stakeholders: such an artifact is also called a *spanning object* (Strode et al., 2012).

Our main contributions in this paper are fourfold:
- C1: We present the ScrumScale scalability model, which exploits the ScrumScale conceptual scalability framework (Brataas and Fægri, 2017, Brataas and Herbst et al., 2017 and Brataas et al., 2020). We also show how we evaluated its usefulness with TietoEVRY.
- C2: We report, in full detail, the ScrumScale conceptual framework used as the foundation for the scalability education in the organization and the ScrumScale model's development. The educational value of the ScrumScale framework was identified as one of the crucial benefits during this study by TietoEVRY.

- C3: We further validate the first steps of the ScrumScale method. This is necessary to ensure that the ScrumScale model is successfully employed and is a valuable result on its own, as the ScrumScale method may be context-dependent.
- C4: We report the lessons learned on how the ScrumScale model was developed and evaluated using the design science process. This provides valuable insights for other organizations to establish similar artifacts and tailor them to their contexts.

This paper is structured as follows: In Section 2, we explore the state of the art in agile scalability requirements elicitation and agile architecture. This section also briefly describes the ScrumScale method. Section 3 presents the scalability concepts in the ScrumScale framework. The study context and business environment for this open banking case study are outlined in Section 4. Section 5 offers our research approach for this case study. The ScrumScale model is described in Section 6. Evaluation during testing is described in Section 7. We separate further evaluation between two different phases. The evaluation during the method development is presented in Section 8. Interviews with the open banking stakeholders in TietoEVRY are discussed in Section 9. A discussion of the contributions of this paper and the threats to validity are found in Section 10. Conclusions and further work are offered in Section 11.

# 2   State of the Art

As a basis for our work, we first briefly describe the state of the art in agile scalability requirements elicitation and, afterward, in agile architecting. We also introduce the ScrumScale method. In the following Section 3, we describe what we mean by scalability using the ScrumScale framework.

## 2.1   Agile Scalability Requirements Engineering

Based on early experiences in ScrumScale, scalability requirements typically suffer from the "curse of the commons," where many functions and customers benefit, but without anyone having clear ownership of the requirement (Brataas and Fægri, 2017). As a result, scalability requirements commonly end up being implicit and unspecified, to the degree that they rarely need revision, with scalability requirements like "the software should scale vertically and horizontally" (ibid). Hence, developers must infer scalability requirements that may not be consistent with other stakeholders' (implicit) scalability requirements. Testers must guess scalability requirements by reading documentation, as architects and developers are often unavailable because they are working on new projects. In the end, actual scalability requirements often surface when the system is in operation, leading to costly and time-consuming redesign.

Neglect of non-functional requirements (NFRs) is common in agile software development (ASD) (Inayat et al., 2015, Ramesh et al., 2010, Behutiye et al., 2020, Medeiros et al., 2020). According to Behutiye et al., 2020, the primary challenge is the limited ability of ASD methods to handle NFRs because of insufficiencies related to eliciting, analyzing, modeling, documenting, and managing NFRs. Further challenges are time constraints because of short iteration cycles, frequent deployments, limitations in testing NFRs, neglect of NFRs, lack of an overall picture of NFRs, and overlooking NFRs by customers.

Alsaquaf et al., 2019, interviewed 17 practitioners about quality challenges in large-scale distributed ASD. Examples of challenges are late detection of quality requirements' (NFRs') infeasibility, hidden assumptions in inter-team collaboration, uneven team maturity on knowledge related to architecturally significant requirements, inadequate non-functional requirement test specification, a lengthy NFR acceptance checklist, sporadic adherence to quality guidelines, overlooked sources of NFRs, lack of NFR visibility, ambiguous NFR communication processes, unclear conceptual definition of NFRs, and confusion regarding NFR specification approaches. As mechanisms behind these challenges, they list implementing NFRs based on unstated assumptions, sub-optimal priority assignment, focusing on a specific viewpoint or component, and losing sight of the big picture. They also describe how customers are not interested in internal NFRs, how legacy architectural decisions complicate the implementation of NFRs on new systems, and the challenges of moving from a waterfall mindset to an agile mindset. In Alasquaf's paper, the challenges were mitigated by using assumption wiki-pages, multiple product backlogs, and automated monitoring tools; reserving parts of the sprint for essential quality requirements; doing sprint allocation based on multiple product backlogs; and establishing a preparation team, a components team, and a quality requirements specialist team.

In a literature review, Ghanbari and Variainen, 2018, found that 26% of the primary studies report that omission of quality practices occurs during requirements analysis specifications. Requirements elicitation is traditionally the first stage in building an understanding of the problem that the software is required to solve (Bourque and Fairley, 2014). Wong et al., 2017, identifies the following six core activities of requirements elicitation: (1) acquire knowledge of the domain, (2) determine the sources of requirements, (3) define the appropriate elicitation technique, (4) identify the requirements of these sources, (5) document, and (6) refine the requirements. For scalability requirements, these activities can be complex. In this paper, we define an elicitation technique for scalability requirements. We also conduct the other five activities.

The principal requirement elicitation techniques are interviews, scenarios, prototypes, facilitated meetings, observations, and user stories. According to Pacheco et al., 2018, these techniques are still the most common. Pacheco et al., 2018, and Aldave et al. 2019 describe more techniques like collaboration techniques (focus groups, workshops, and brainstorming), contextual techniques (combining unstructured interviews and prototyping), gamification, and storyboards. Using the ScrumScale model as a spanning object, we complement these elicitation techniques.

Scalability is the ability of software to increase its capacity by consuming more (hardware) resources (Brataas and Herbst et al., 2017). This is in line with the scalability definition proposed by Lehrig et al. in a structured literature review of scalability definitions in cloud computing: *"Scalability is the ability of a cloud layer to increase its capacity by expanding its quantity of consumed lower-layer services"* (Lehrig et al., 2015). Furthermore, our definition is also consistent with the principles described by Jogalekar and Woodside, 2000. As described in Section 3, scalability builds on capacity and performance but focuses on workload growth. Software performance engineering (SPE) is a well-established field (Smith and Williams, 2001), in which the focus traditionally has been on building models of software systems to evaluate design trade-offs. The overall aim is to avoid the "fix-it-later" approach, where functional requirements get (all) priority and performance problems are ignored until late in development. The "fix-it-later" approach increases technical debt and results in costly and time-consuming performance tuning and subsequent performance refactoring. Existing SPE performance and scalability models, tools, methods, and guidelines are valuable but are usually time-consuming and require considerable manual work from skilled personnel (Becker et al., 2017). Measurements during performance testing represents another prominent technique for improving the performance of a system. However, performance testing is not applicable if you do not know the workload but would like to estimate the workload, as in our case.

Wohlrab et al., 2014, describes the PROPRE method for performance requirements engineering. This paper describes six interesting criteria for performance requirement engineering methods: (1) easy for all relevant stakeholders to understand, (2) include stakeholders in the process and encourage discussions, (3) help to focus on essential requirements to reduce time and effort, (4) result in a good basis for requirements' specification and the creation of test scenarios, (5) be suitable in a distributed business scenario, in which it is hard to schedule meetings where all stakeholders can participate, and (6) be applicable under time constraints. Wohlrab et al. did not find any performance requirement methods fulfilling these requirements. They propose the PROPRE model based on abstract feature models and feature sheets. The feature model in PROPRE presents all operations in the system annotated with a performance metric. In the feature sheet, they rank the importance and difficulty of all features. By difficulty, they mean the development and test effort needed to improve and measure the feature's performance. In Scrum, the feature model and the importance are not required, as the priority of each user story already takes care of them. The difficulty is also handled by the estimation of effort in Scrum. Each requirement in PROPRE is presented as an extension of a so-called Volere shell. This could serve as an inspiration for future extensions of the ScrumScale model. The PROPRE approach focuses on performance, which is related to but not the same as scalability. In conclusion, PROPRE is not agile but should be considered in future extensions of the ScrumScale model.

The ScrumScale method describes, as illustrated in Figure 2, how to work with scalability requirements during agile development (Brataas et al., 2020). The ScrumScale method consists of Sprint 0 with three necessary preparations before regular development sprints, followed by similar steps 1, 2, 3, 4, which are done at each sprint iteration. Sprint 0 starts with defining the (functional) user stories in Step A. Afterward; step B performs a scalability triage, an informal and quick meeting during which one or more scalability experts discuss scalability risks associated with these (functional) user stories. The concept of triage comes from emergency medicine, in which a doctor quickly determines whether a person requires immediate treatment or can wait. If a user story has scalability risks, we continue to work with this user story in the rest of the ScrumScale method; otherwise, we do not need to consider this user story anymore in the ScrumScale method (Hanssen et al., 2019). With new information, this decision may, of course, be revisited. The primary sign of a risky user story is an alarming relation between work, load, and quality thresholds for one or more user stories, e.g., high work, high load, and tough quality thresholds.

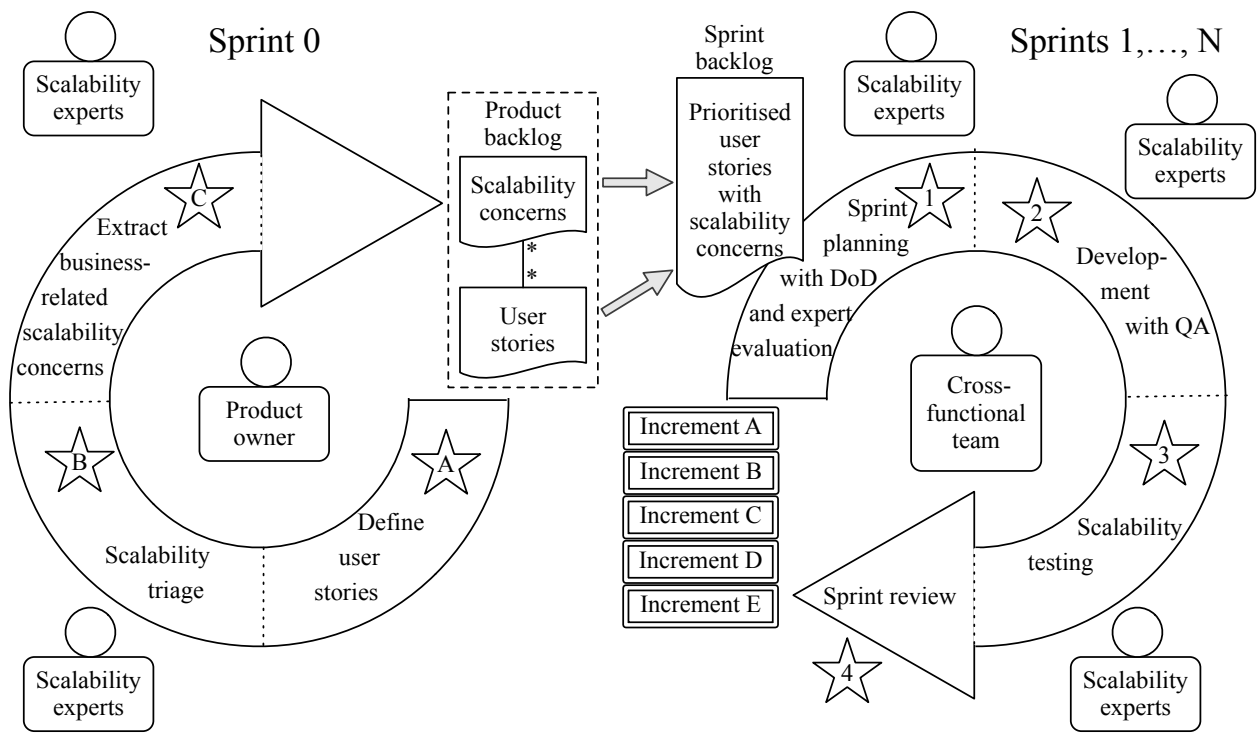

Figure 2: The ScrumScale method (Brataas et al., 2020).

In the ScrumScale method, scalability experts span different competencies and, accordingly, different persons and existing roles, like software architect and performance tester. Step C in the preparatory Sprint 0 is to extract business-related scalability requirements, where 12 different issues are questioned. These issues tell us to find information reflecting the ScrumScale conceptual framework: work, load, quality thresholds, plus supplementary information related to the planning horizon, system boundaries, and consistency. When Sprint 0 is completed, the initial user stories in the product backlog are complemented by scalability requirements. In later sprints, 1, 2, 3, …, N, the scalability requirements are further refined, tested, and potentially also relaxed as details of the system under development grow.

In summary, broader work on NFRs in ASD confirms that scalability requirements elicitation is often neglected and faces several challenges. When working with scalability requirements, we aim for a pragmatic middle ground between 'laissez-faire' using fix-it-later and over-engineering with extensive modeling. We base our work on the ScrumScale method (Brataas et al., 2020), which we have developed earlier. This allows for sharing insights into scalability using a lightweight method based on a conceptual scalability elicitation framework. This framework assists stakeholders in their collaboration on scalability engineering.

## 2.2   Agile Architecture and Coordination

Software architecture is the technical solution that results from a tradeoff among different system qualities. Such qualities or NFRs can be interdependent with functional requirements. In fact, from requirements analysis, architects distill architecturally significant requirements, which drive the implementation of the technical solution. Architecture is also considered one of the major coordination mechanisms across teams in a large company (Herbsleb, 2007).

Traditionally, architecture has mostly been an upfront activity, in which the software blueprint has been designed before the implementation begins. Nevertheless, with the need to constantly adapt to market needs and the advent of ASD, software architects have tried to make the process of defining architecture more iterative. This, however, is very dependent on the input from the requirements analysis.

Basic agile methods do not usually provide much support to methodically grow a product architecture. In contrast, the traditional architecting methods are based chiefly on massive upfront design, which is more in line with a waterfall approach. Current research on agile architecting is trying to find a sweet spot by combining agile and architecture to provide hybrid techniques.

Agile architecting (also known as continuous architecting) is an essential yet under-researched topic. It is, in fact, not yet clear from the scientific literature, as well as practical surveys, how to support an architecture that adjusts to changing requirements and uncertainty (Mårtensson et al., 2019). We, therefore, aimed to follow this line of research with an in-depth case study focusing on scalability.

Bellomo et al., 2014, describes agile architecting in terms of the architectural process: *"An Agile way to define an architecture, using an iterative lifecycle, allowing the architectural design to tactically evolve over time, as the problem and the constraints are better understood."* The paper suggests an iterative process for architecture parallel to the iterations related to requirements and implementation and is synchronized via integration points. We follow such proposed high-level guidelines when designing our approach.

Poort and van Vliet, 2011, argue that to make architecting more agile, architectural concerns (e.g., scalability) should be collected via stakeholders. Then the architectural backlog can be re-prioritized iteratively, according to additional information elicited (in a lightweight manner) during the requirement and implementation activities. Scalability is one such architectural concern, but the paper does not explicitly address it. We, therefore, take inspiration from such a study to design our approach.

Architecture risk management is discussed by Waterman et al., 2015. According to Waterman et al., addressing risks conflicts with responding to change. Understanding how to mitigate such conflicts is, therefore, important. ATAM (Bass et al., 2012) is an approach for handling many different non-functional requirements. However, ATAM is an extensive upfront analysis relying on an existing architecture. Our goal has been to find a more continuous and lightweight approach.

Another relevant perspective is agile architecture communication. In our previous work (Martini and Bosch 2016a), we found that it is crucial to pay attention to agile communication between architects and development teams (including testers), which should happen iteratively (in both directions). In particular, architects must clearly and continuously communicate the importance of architecturally significant requirements (in our example, concerning scalability). At the same time, they must have feedback on the status of the system to understand whether the architectural requirements are at risk. However, this paper does not dive into the specific problem of communication related to scalability.

An essential aspect of managing NFRs such as scalability is the coordination across several stakeholders. According to available theories of coordination in agile software development (Strode, et al., 2012), agile projects must coordinate via spanning activities, spanning objects, and spanning roles. Current literature covered mostly small and co-located projects, where NFRs are mostly locally taken care of by agile teams. However, in large projects, coordination across teams to manage cross-cutting concerns, such as scalability, is a challenge (Bick et al., 2018) and requires spanning activities, objects, and roles. We take inspiration from these coordination mechanisms to develop our approach.

While the ScrumScale method (Brataas et al., 2020) can be considered a spanning activity, our primary goal is to develop a spanning object, the ScrumScale model, to coordinate the input across the different stakeholders involved analysis of scalability. The ScrumScale method presents a new role, the scalability expert that supports the team in sharing the relevant information across the stakeholders and making the right decisions to meet scalability requirements or concerns. Such role can be considered a spanning role, making use of the spanning object (the ScrumScale model) as a tool. Besides collecting and reporting the case, the researchers coached and supported the organization with conceptual knowledge from literature and shared scalability tools, concepts, and relevant methods using the ShareScale model, the ScrumScale framework, and the ScrumScale method.

In conclusion, several studies advocate the need to manage the risks posed by architectural decisions in the presence of uncertainty and related to NFRs. Such studies propose an agile and lightweight approach, which this study embraced. However, none of the studies report a concrete approach to specifically manage scalability in such a way, especially with a lightweight spanning object (e.g., a model), which motivates our in-depth study. Therefore we have considered the high-level guidelines reported in the available studies to develop our artifact (the ScrumScale model).

# 3    Conceptual Scalability Framework

In the ScrumScale project, several practitioners commented that the scalability concepts that we proposed were obvious (Brataas and Fægri, 2017). On the other hand, these concepts were not used in practice when eliciting scalability requirements. Project after project failed to deliver sufficient levels of scalability, partly because of poorly elicited scalability requirements. We concluded that the complexity in managing scalability was probably not buried in the individual concepts but, instead, in the lack of a systematic approach to applying *the set of concepts* relevant for scalability. This set of scalability concepts is the ScrumScale framework. This framework is vital for the education of practitioners, the understanding of the ScrumScale method, and the understanding of the ScrumScale model. This section introduces the concepts in the conceptual scalability framework one by one before introducing the complete scalability framework.

The conceptual scalability framework gathers concepts that form the foundation of our work on scalability requirements. Upon this framework, we build the scalability model described in Section 6. This section presents the framework in a clear, self-contained, consistent, complete, reader-friendly way. This presentation is based on the written educational material given to TietoEVRY. It is also a textual version of what was orally presented to the stakeholders in the open banking case study. In (Brataas and Fægri, 2017), we presented an initial version of our conceptual scalability framework, which was further elaborated in (Brataas and Herbst et al., 2017) and, briefly, also in the ScrumScale method (Brataas et al., 2020).

## 3.1   System

A *system* contains *services* and *resources*. By *services*, we mean *software* services. There are many types of *resources*. Typical *cloud resources* are SaaS (software as a service) or IaaS (infrastructure as a service). Active *hardware resources* are CPUs (for processing), disks and memory (for storage), and network equipment (for communication). Passive *software resources* like locks, buffers, and semaphores are typically associated with handling consistency in storage and often represent scalability limitations.

The *system boundary* separates services and resources within the system and services and resources outside the system. Services and resources outside the system boundary are not included when we measure response times for the system. In an *open system*, the users are outside the system and may, therefore, vary between zero users and many users. A web-based system is a typical example of an open system. In a *closed system*, the users are inside the system, together with services and resources. The number of users is therefore constant. This is usually the case when we test a system with an artificial load. There may be several *types of users*: the public, bureaucrats, administrators, specialists with heavy operations, etc. Note that these definitions of open and closed systems in a performance/scalability context are stricter than the definitions of open and closed in a business context related to open banking as described in Section 1.

## 3.2   Operations, Work, Load, and Workload

As illustrated in Figure 3, *operations* represent different ways of interacting with a *service*, e.g., functions, calls, transactions, queries, jobs, and user stories. These operations will be related to processing, storage, or communication. These operations will typically use some *work objects* that are internal to the system, like documents, drawings, images, movies, etc. In our bank domain, we have more complex work objects like accounts and payment history. Especially if these objects vary in size, it is essential to characterize them with *work parameters*, e.g., number of accounts and length of payment history. Work parameters represent information characterizing what is done each time we invoke an operation.

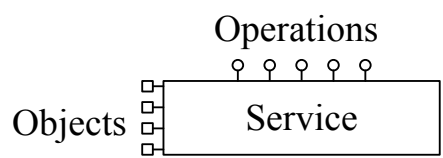

Figure 3 A service has operations and may operate on objects.

*Load* specifies the number of requests for an operation within a given period. For open systems, the load is characterized by intensity, e.g., transactions per second (TPS). For a closed system, the load is specified by the number of customers in the system and average think times. The think time is the time between receiving the result of one request from the (computer) system and sending a new request (typically by a human to the computer system). The *workload* is the product of work and load. We do not usually compute this product.

Commonly, the concept of workload intensity is used instead of load (Smith and Williams, 2001). Smith and Williams do not explicitly mention work but uses the concept of "work units" to specify the relative amount of work for each invocation of an operation—for example, linearly ranging from 1 for a light operation to 5 for a heavy operation. We shall later see that the explicit introduction of work makes our conceptual model more precise when assessing scalability.

We are interested in the highest load, i.e., the load during the busiest hour, during the busiest week, during the busiest month, and in the busiest year. To estimate this load, the concept of burstiness is helpful. In this article, burstiness is the ratio between the maximum load and the average load. Assume a period $T$ is divided into several shorter periods $t$. Let $x$ be the average load in period $T$ and let $y$ be the *highest* average load in one of the $t$ periods. Then burstiness $b = y / x$. Assume the average load for a typical day is 100 transactions per hour. A burstiness of 5 means that for the busiest hour, the load is 500 transactions per hour. If a service is used with a flat rate for five active hours a day and not used in the remaining hours, this corresponds to a burstiness of 4.8 (24/5 = 4.8).

## 3.3   Quality Metrics and Thresholds

A *quality metric*, in our context, describes how to measure the quality of service. Response times are typical examples, e.g., average response times and 90-percentile response times. *Quality thresholds* are the actual number for each quality metric and distinguish between acceptable and not acceptable quality of service. Often, all operations share the same quality metric. Still, the *quality threshold* may differ between the operations, e.g., some operations have a two-second 90-percentile response time, while others have a five-second 90-percentile response time. *The critical operations* are the operations in which there is a risk of not satisfying the quality thresholds.

## 3.4   Capacity

At an overall level, *capacity* is the maximum workload a system can handle while still satisfying quality metrics and quality thresholds. Because workload means both work and load, it is more straightforward if we vary either work or load and keep the other fixed. When capacity is measured in terms of work, we must specify which work parameters we mean, as there are commonly several work parameters (Brataas et al., 2018b). In this article, we focus on capacity in terms of load. As a result, all work parameters, quality thresholds, and deployment configurations are fixed. This is illustrated in Figure 4, which shows how quality degrades as we increase the load on a system. Initially, quality—typically measured with response times—is reasonably stable. As we approach full utilization of the system, the quality degrades quickly. The system's capacity is the highest load that fulfills the quality thresholds of all operations in the system. Therefore, if the quality metric for all operations is the 90-percentile response time (as in our case study in this article), for all these operations, no more than 10 percent of the requests can violate the threshold. To measure the capacity, you must perform several measurements, each time with a stable workload. The workload when quality thresholds for all the operations are just barely obeyed is the capacity.

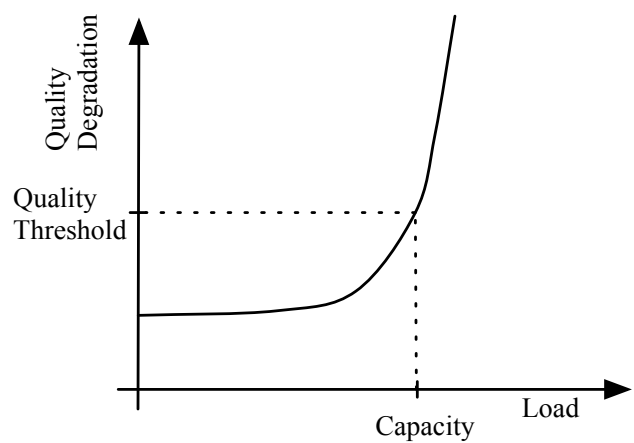

Figure 4: The capacity of a system (adapted from Rygg et al., 2013).

## 3.5  Scalability

*Scalability* is defined as a system's ability to increase its capacity by consuming more resources (Brataas and Herbst et al., 2017). Figure 5 illustrates the scalability of a typical system. This figure shows the relation between the cost of the resources involved and the capacity of the system. In addition to the cost of cloud and hardware resources, software license costs may be relevant. In practice (Rygg et al., 2013, Brataas and Herbst et al., 2017), the shape of the graph, of course, varies from case to case. Capacity may also be measured in terms of work (Brataas et al., 2018b).

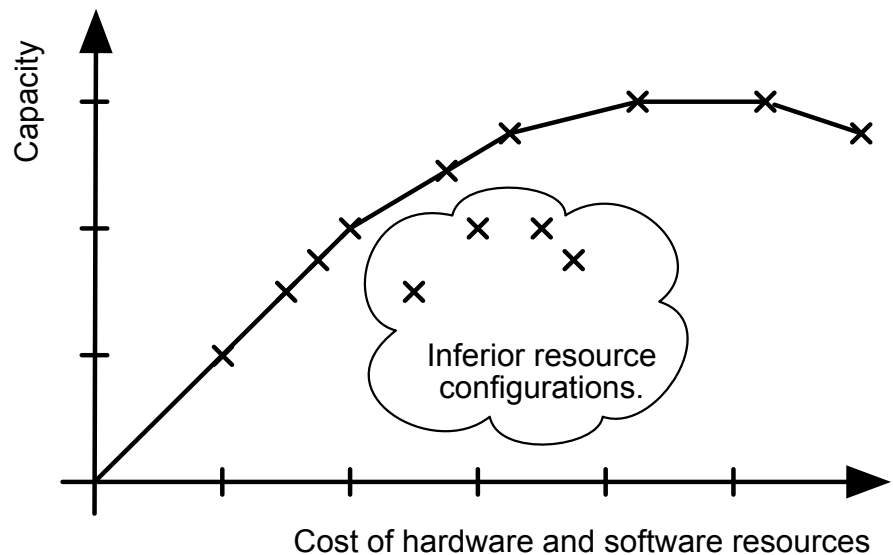

Figure 5: Scalability as the relation between cost of resources and overall capacity (Brataas et al., 2020).

The resulting relation in Figure 5, between the amount of resources and the resulting overall capacity, is, for humans, termed the Ringelmann effect (Kravitz and Martin, 1986). The Ringelmann effect (for humans) is caused by either laziness or coordination overhead. Only the latter applies to computers. Scalability testing establishes graphs like the one in Figure 5, whereas performance testing focuses more on one capacity value for one given amount of resources, like in Figure 4.

Comparing our scalability framework to Jogalekar and Woodside, 2000, reveal similarities and differences. Both Jogalekar et al. and we use the cost of hardware and software resources for describing the size of a system. However, while quality thresholds are variable in Jogalekar et al., we primarily keep the quality thresholds constant during scaling. We use quality thresholds to find the capacity, whereas Jogalekar et al. uses throughput, response times, and values of the response times as inputs to the scalability analysis. Jogalekar and Woodside leave out the concept of work, which can also be used to find capacity in our scalability framework (Brataas et al., 2018b). In our scalability framework, capacity can also be found be varying quality thresholds, estimating the sensitivity to changes in quality thresholds (Brataas and Herbst et al., 2017).

Scalability problems are often deeply rooted in the system architecture and may be hard to tune away, e.g., a centralized SQL database may be more limiting than a non-SQL document-based database with relaxed consistency requirements (Brataas et al., 2020). It is wise to spot such *scalability anti-patterns* early. A *scalability pattern*, on the other hand, distills well-proven expert knowledge and is helpful during design. The concepts used to describe scalability are summarized in Figure 6. In contrast to Figure 6, Figure 3 has a narrower focus on services and shows (work) objects. Figure 6 illustrates how (work) objects have work (parameters).

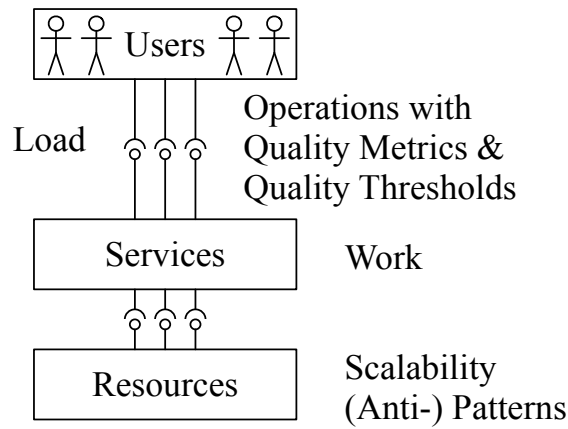

Figure 6: Collection of scalability concepts: The ScrumScale framework
(adapted from Brataas and Fægri, 2017)

## 4   Study Context

This section describes the context (or "the environment" as defined in design research) in which the study has taken place. It is essential to understand the constraints posed by such a context and determine the business needs we want to satisfy in our design science process when designing the new artifact.

In November 2015, the European Union passed the Revised Directive on Payment Services (PSD2), which requires financial institutions to open up their services to authorized third-party providers (TPPs) through APIs (application programming interfaces). The goal of PSD2 is to increase pan-European competition, participation in the financial industry (including non-banks), and encourage innovation. A typical example of a third party's service through such open interfaces is an overview of an end user's balances and recent transactions across various banks. By March 14th, 2019, financial institutions were to offer APIs for external testing, and by September 14th of the same year, compliance with the directive became mandatory.

For TietoEVRY, a European software vendor offering solutions covering most areas within the banking sector, PSD2 implied technical updates to realize the desired APIs. However, more importantly, it implied that new and unknown traffic patterns were to be handled by a mixture of legacy and newer systems, some of them external to TietoEVRY. To address these requirements, TietoEVRY launched a project, Open Banking. The concept of open banking is a generalization of the PSD2 requirements, as it may encompass open access to any financial services API, not just those stipulated by PSD2. The Open Banking project aimed to build an access platform for TPP request routing, call authorization, and traffic control.

Some of the challenges identified include:

- PSD2 states only in general terms what performance or scalability is expected from an API. It says, for example, that performance should be the same through the API as it is directly to the bank's users, in terms of both response times and the freshness of the data. Performance shall be monitored and reported to the relevant financial supervisory authority so that compliance can be verified.
- If an API is not in place for an operation mandated by PSD2, a TPP shall be allowed to do screen scraping, i.e., use an available end-user client and pick up data directly from the user interface. This raises several issues, such as how to limit the information visible to the TPP.
- TietoEVRY chose to realize the Open Banking solution by implementing standards defined by the Berlin Group (https://www.berlin-group.org/). This specification turned out to be a moving target, causing the re-writing of platform software. The specification also contains many options and degrees of freedom, so communication and clarification with available TPPs must be done.

On the other hand, PSD2 also puts limitations on what TPPs are allowed to do:

- A TPP can access data only for individuals who have explicitly authorized such access, and such authorization shall be given per financial institution.
- There is also a limit on the number of requests that a TPP can issue per customer per day (currently four). This does not include requests that the customer initiates.

The Open Banking access platform funnels TPP requests to back-end systems, such as core banking and payment applications. These are complex systems, so most requests will probably have high work parameters (see Section 3). An effort was made to optimize relevant back-end system operations for open banking. It is also possible to separate the open banking workload from the remaining workload.

When assessing the needed capacity and scalability of the APIs offered through the Open Banking platform, TietoEVRY had to estimate future usage patterns. Parameters here are the number of TPPs, the number of banks providing APIs through the platform, and the degree to which bank users will be interested in and use the TPPs' services, all of which are largely unknown. Furthermore, some of this new traffic will simply replace existing traffic, making the analysis challenging. Therefore, the approach chosen was to work out several workload scenarios based on different assumptions, as detailed later in Section 6.

Once the scenarios were worked out, the natural next step was to test the system to see whether the performance objectives would be met. This effort also met with several challenges: It was impossible to test in production with high workload (Section 8), as this would disturb regular traffic. The available test environment was not identical to production, and some services were just mocked, so results would have to be interpreted. The value chain in some cases included external customer systems, making it hard to control or emulate a realistic environment.

Thus, reasoning about and securing adequate scalability of solutions offering open banking APIs face several elements of **uncertainty**:

- The **requirements** in scalability are uncertain because we know little about the plans of third parties and customer banks, as well as the usage patterns of end users adopting the new services.
- The **implementation** was, in this case, uncertain due to changing and open-ended specifications.
- The **scope** in terms of the involved value chain was complex and might contain systems outside the control of the implementing organization.
- The **testing** produced uncertain results due to the unavailability of a realistic test environment, complex value chains, and, usually, also a lack of comprehensive test data and workload. In addition, testing should be done early, that is, on a partially developed system, further complicating the extrapolation of test results to the final production system. See (Brataas et al., 2018a) for more details on the challenges of testing in ASD.

These and other uncertainties seem to be increasingly common in today's IT ecosystems, which often exhibit such characteristics as long and complex value chains, many system interconnections, open interfaces with limited control of demand, rapid changes in traffic and usage patterns due to sudden shifts of various kinds, large user populations, and distributed control. In this "age of uncertainty," the ScrumScale method offers a light-weight approach to gather as much information as is available at a particular stage in a structured manner, making the best use of it to point the solution in a useful direction and solicit any information that is still missing to arrive at a good result.

# 5   Method

As a guide to ensuring a good choice and application of research method, we have consulted a set of quality criteria for qualitative studies as described by (Dybå and Dingsøyr, 2008) to ensure that our research is 1) rigorous (the research method is appropriate and has been applied thoroughly), 2) credible (the findings are well-presented and meaningful), and 3) relevant (the results are helpful to the software industry and the research community). In particular, in the next section, we describe why and how we thoroughly followed the Design Science Research (DSR) approach (Hevner and Chatterjee, 2010), which encompasses methods and guidelines to ensure the production of credible and relevant findings.

## 5.1   Design Science Research Approach

To answer our research questions, we chose to follow a DSR approach. Our goal was to *Develop* and *Evaluate* an artifact that would be used in practice by the studied company, and that could be used in other contexts.

Our DSR is based on the framework outlined in Figure 7, extracted, and simplified from Hevner and Chatterjee, 2010. The *Knowledge Base* and *Environment* (respectively reported in Section 2 and 3) provide our Design Science Process, respectively, with the *Business Needs* from the context and the *Applicable Knowledge* from existing literature. The artifact is then developed and evaluated multiple times through several refinement stages. Using the design science process and ensuring a rigorous research method to develop and assess the final artifact also allows us to systematically report our findings, which increases the transferability of our results to other contexts.

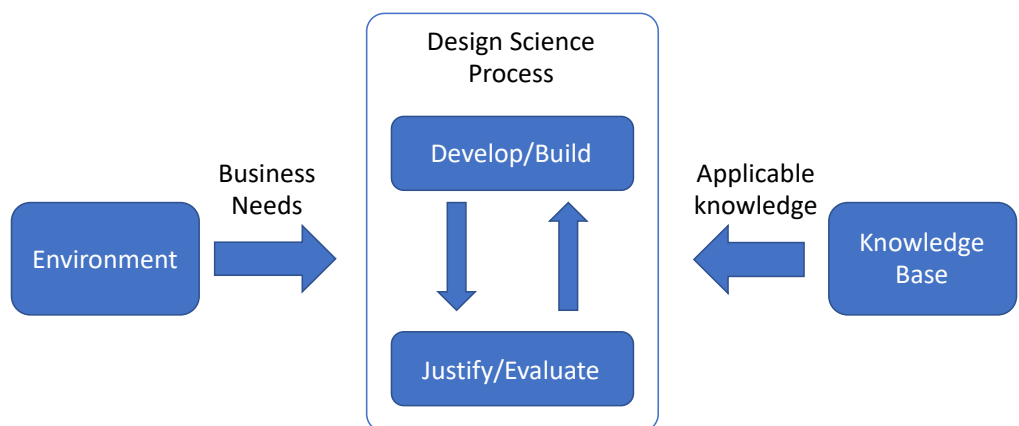

Figure 7: Design science process (extracted from Hevner and Chatterjee, 2010)

Our implementation of the DSR followed the guidelines in Hevner and Chatterjee, 2010 are as explained below.

## 5.2   Design Science Overall Process

The output of our design research consists of the ScrumScale model, a spanning object used to analyze scalability requirements in combination with a lightweight process (the ScrumScale method) by scalability experts (spanning roles). The design process prescribes two main phases: *developing* and *evaluating* such artifacts. These two phases were broken down into six main steps, which are outlined in Table 1.

The first four steps are related to the **development** of the model and were:
1) *Project kickoff:* Get an overall understanding of the system being affected by the PSD2 regulations and of its scalability challenges, as well as discussing the matter of which stakeholders we should speak to to get the scalability requirements.
2) *Eliciting structure for the ScrumScale model*: Together with the stakeholders, we prepared a spreadsheet and identified the critical information and how to visualize it.
3) *Subsystem scalability analysis:* We prioritized the information that should have been in the model (e.g., which operations, which parameters, etc.). We talked with the different stakeholders from four subsystems to understand which operations should be included in the model.

4) *Scalability parameters checking:* We refined the model with input from the stakeholders and discussed with management to receive approval to use and evaluate our model in practice.

Note that, for this kind of artifact, the development and refinement of the model also implies that the artifact was **applied** in practice and evaluated by the practitioners continuously. The continuous application and refinement of the model are at the core of the design science research methodology. It helps to receive feedback from the users of the methods. To better illustrate the continuous development and refinement of the model, we report In Table 1 the correspondence between the artifact refinement steps and the software development project sprints 0-N.

To finally **evaluate** the artifact (besides the continuous application and evaluation mentioned above), we proceeded with the following two steps:

5) *Preparation of scalability testing:* We worked closely with the testing stakeholders to prepare the open banking application's performance to be tested in a production environment. This preparation by the stakeholders provided valuable evaluation of the ScrumScale model. Because of late project constraints, which caused limited hardware and personnel resources, limited performance testing, but no real scalability testing was done.

6) *Evaluation of process and results with stakeholders*: We collected feedback from the key stakeholders about how the lightweight process and the model were used in practice. We collected the benefits and challenges of using the artifacts. We defined this as the limit of the study. This is in line with Hevner and Chatterjee, 2010. They report that descriptive methods of evaluation can be used for especially innovative artifacts for which other forms of evaluation may not be feasible (as was the case for this case study, since it was not possible to perform scalability testing because of project constraints). Further investigation in later development projects could be to evaluate the feasibility and value of the ScrumScale model more rigorously.

Table 1 also reports which participants were involved in the meetings, number of meetings required for each step, and meeting hours and person hours to carry out each step. The *Output* column shows the types of findings obtained concerning the phase of the design science process (namely, the *refinement* of the artifact, the *scalability testing*, and the final assessment of the *benefit and challenges*).

In Table 1, the person hours count only when the TietoEVRY—and not the researchers—spent in the meetings. It would have been hard to distinguish between the research time required to do productive work in this case study and the research time used to develop the approach. For TietoEVRY, it was easier because they worked only with the case study during the described meetings. With a mature approach, the researchers would not be involved. Instead, the company would need some more hours for the spanning role. On the other hand, the number and length of the meetings could have been streamlined. Therefore, the number of TietoEVRY person hours indicates the effort involved in this case study.

In this section, we report, for each phase in the design-science process in Table 1, how we carried out our data collection activities.

### 5.2.1   Environment and business needs

The artifacts were developed together with TietoEVRY, which was facing a challenge related to implementing operations with changing and uncertain scalability requirements for the open banking framework. This context gave us the relevant business goals to justify our research. Such business goals were collected with stakeholders at the collaborating company, including the first authors of this paper. See Section 3 for the environment description.

### 5.2.2    Development of the artifacts

The artifacts were developed and used in practice during a real-life case in a large software company. We systematically collected insights via a case study including 36 meetings, for a total of 35.5 hours with the following key stakeholders (see Table 1 for more details):

A.  Two solution architects, responsible for the scalability of open banking and the Core subsystem
B.  Head of non-functional testing in TietoEVRY
C.  Responsible for the system development process and fourth author of this paper
D.  Chief technical architect for the Payment subsystem and security
E.  Chief architect and team responsible for the Core subsystem, the most critical open banking subsystem
F.  Development manager for another subsystem
G.  Enterprise architect across many financial systems in TietoEVRY
H.  Responsible for performance testing in the open banking project
I.  Project manager for open banking
J.  Architect for open banking

## 5.3   Data Collection

Most of the meetings were attended by two researchers and were recorded. The researchers produced a summary of each meeting, reporting the information that was important to the research process. Given the nature of the process, meetings often included practical information related to the context, but not necessarily need to be reported in relation to the research process. Such information was filtered out.

### 5.3.1   Scalability testing

In the first part of the evaluation, we prepared for the scalability testing, which was partially executed. We discussed several aspects related to scalability testing:
- Should subsystems be included when testing open banking?
- Which tests are required to declare compliance to the PDS2 directive, and how do we best exploit the test results we obtained?
- Which test rigs are available, and which test rigs should be used?
- Which components do these test rigs consist of, and which ones are likely to become bottlenecks?
- Should the test be performed in the test environment or in the production environment?
- How representative will these tests be?
- How should the load be generated in the different environments?
- Which quality thresholds should be used when testing?
- Which think times should be used when testing?
- When will the dynamic mock be available? This was required to emulate part of the production environment in a more synthetic setting.
- How do we understand the production test results?

Table 1. Phases in the open banking case study.

| Dates | Purpose | Stakeholders | #meetings | Meeting hours (person hours) | Output |
|---|---|---|---|---|---|
| 2018 May | Project kickoff (Sprint 0) | Head of NF testing, solution architect, sys. dev. responsible, two researchers | 1 | 1 (3) | |

| 2018 May – July | Structure the ScrumScale model (Sprint 0) | Head of NF testing, solution architect, sys. dev. responsible, two researchers | 6 | 8 (11.5) | Artifact v. 1 |
|---|---|---|---|---|---|
| 2018 August | Subsystem scalability analysis (Sprint D) | Chief technical architect, dev. mgm., enterprise architect, solution architect, one researcher | 4 | 3.5 (3.5) | refinement |
| 2018 September – 2019 March | Scalability parameters checking (Sprint F-J) | Project Manager, chief technical architect, solution architects, performance resp., sys. dev. responsible, two researchers | 9 | 11.5 (17.5) | Artifact v. 2 refinement Artifact v. N |
| 2019 March – October | Preparation for scalability testing (Sprint K-N) | Performance resp., sys. dev. responsible, architect, two researchers | 9 | 8 (14.5) | Scalability Testing |
| 2019 October – November | Evaluation | Project Manager, solution architect, development mgm., performance resp. sys. dev. responsible, head of NF testing, two researchers | 7 | 3.5 (5) | Benefits & challenges |
| SUM | | | 36 | 33.5 (55) | |

Many of these discussions go far beyond the focus of this paper, mainly because limited performance testing was actually done. During these meetings, we also understood several aspects of open banking, such as the role of the Berlin Group and the relation between open banking and PSD2.

### 5.3.2   Evaluation of the artifacts

During the evaluation, we carried out seven semi-structured interviews with key personnel related to the scalability of open banking in TietoEVRY. We talked to the head of non-functional testing, the initial solutions architect for open banking, the present architect for open banking, the project manager for open banking, and the person responsible for performance/scalability testing of open banking. Moreover, we spoke to the chief architect and the solution architect in the most important subsystem in open banking, the Core subsystem. These seven interviews lasted between 15 minutes and one hour. Our initial questions were formulated as:

- What are the challenges with handling scalability for open systems?
- How useful was the approach?
- How effective was the approach?
- What ideas do you have for further improvement of the approach?
- For the full benefit of the method, what should be changed in TietoEVRY?

- What is the potential of the ScrumScale method?
- Are there issues with this method that should have been addressed but that it does not handle?

Based on feedback in the first interviews, the questions were further refined. The last interviews with the subsystem representatives focused more on how the ScrumScale method could also have been helpful in this project.

## 5.4   Data Analysis

One researcher analyzed the data, while the codes were checked with another researcher when uncertainty arose. We used a version of the thematic analysis methods described in Clarke et al., 2015. In particular, we used a deductive approach based on our research questions and the study's goal. Then we followed the recommended steps: familiarization with the data, coding the data, searching for themes (in our case, it was more of a mapping exercise), reviewing and defining names for the themes.

In practice, we first defined categories in which we were interested, according to the research questions. Then we coded the data and filled the categories with the codes and quotations while creating new categories if needed. The categories differed concerning which phase of the design science process we were focusing on (development or evaluation), as different phases would give us different insights.

For the **development** phase, the main categories were:
- *Context:* We extracted information concerning the context of this case study. This allowed us to report better our findings and how they could be transferred to other organizations with a similar context.
- *Artifact (the ScrumScale model)*: We wanted to capture specific characteristics of the model that we developed. From the design perspective, we should understand what led the artifact to be successful (or not). In particular, we studied:
  - *Initial structure:* We gathered the characteristics of the initial model used.
  - *Incremental changes*: We identified the parameters in the model that were changed during the development.
- *Artifact (ScrumScale method)*: We used a similar categorization (see the previous point) for the analysis of the method; we wanted to understand what made it successful (or not) and what changed incrementally.
- *Artifact (the ScrumScale framework)*: We were interested in the usefulness of applying the ScrumScale framework that underpins the ScrumScale model. The presentation of the ScrumScale framework was refined during several discussions with TietoEVRY personnel.
- *Challenges:* During the development of the artifacts, we tried to capture the kinds of challenges that hindered such development and that we had to overcome.
- *Lightweight prioritization*: One of the critical business needs was to produce a lightweight method to assess scalability debt. Therefore, we looked for elements that were important to making the process lightweight.
- *Communication and coordination:* One of the crucial aspects of our approach was to develop a spanning object and a spanning activity that would allow different stakeholders to communicate and coordinate. Hence, we paid particular attention to this aspect.

We also focused on understanding how the method and artifact were developed iteratively and what differences were found between different versions of them.

For the **evaluation** phase, we explicitly looked for the two categories, *Benefits* and *Challenges*. We wanted to understand, from each interviewed stakeholder, what the artifacts helped them with and what drawbacks the artifacts had. The following section describes this in more detail.

# 6   The ScrumScale Model

In Brataas and Fægri, 2017, we describe how limited collaboration when defining scalability requirements leads to vague scalability requirements, e.g., "our solutions shall scale vertically and horizontally." Vagueness in scalability requirements implies that they rarely need revision, and they, therefore, remain outside the center of attention. In Brataas and Fægri, 2017, we speculate that misplaced and diffuse ownership of scalability requirements is a fundamental explanation of why scalability requirements are subjected to such ad-hoc treatment. Previously, in Section 2, we introduced our conceptual scalability model. This section will describe how these *concepts* are used to create a practical scalability requirement *model* using a simple spreadsheet. This model was initially developed when we talked to one key scalability expert in TietoEVRY. Later, more stakeholders were involved using snowballing.

We will first describe the scalability triage set that reduced the original ten operations to a much more manageable number of three. Afterward, we introduce the three scenarios and then present the parameters in the scalability requirement elicitation model. We conclude this section by describing the essence of the ScrumScale model. This should assist in the reuse of this scalability requirement model.

## 6.1   Scalability Triage

During a scalability triage meeting scalability in step B, scalability experts consider all operations. Operations with scalability risks are tagged and are considered further in the ScrumScale method. Triage reliability can be increased by combining the viewpoints of several scalability experts, and it will be further improved through the sprints. This initial triage decision in the scalability triage meeting in step B is revised in step 1 for all later sprints, especially with changes in system boundaries or any of the operations (Brataas et al., 2020).

As our starting point in this case study, we found ten open banking operations based on available information. Together with scalability experts in the open banking project, we scored the work, the load, and the quality thresholds for all these ten operations. For work and load, we used the four values of L (low), M (medium), H (high), and VH (very high). Following the SLA from TietoEVRY, all ten operations had the same quality threshold, $n$. Because of non-disclosure, the exact quality threshold is not revealed. The name of seven operations is also not made explicit, also because of non-disclosure.

Afterward, we asked the following question: For which operation is there a risk that the product of work and load will not fulfill the quality thresholds? There are no exact criteria for finding the critical operations. The scalability experts make a joint, holistic evaluation when they see the scoring for all operations. However, an operation with high work, high load, and severe quality thresholds would be a critical operation. Similarly, an operation with low work, low load, and loose quality thresholds would not be a critical operation. The discussions will be more challenging for operations with high work, medium load, and medium quality thresholds.

We discussed and found three operations in which this risk was present: payment, balance, and transaction history. This reduction from ten to three operations was crucial for the further success of the scalability elicitation process. Otherwise, the fix-it-later approach and neglect of scalability requirements as described in Section 2.1 would be very tempting, simply because the amount of work involved in further scalability requirement elicitation would seem overwhelming. This reduction is the scalability triage step (step B) in the ScrumScale method, as illustrated in Table 2. In Table 2, operation 3 had very high work but low load. After an expert evaluation, we found that this operation was not critical. For operation 10, we did not manage to set the work, but after an expert evaluation, we found that this operation was not critical either. The set of three critical operations, initially identified in Sprint 0, did not change when we revisited this decision in Sprints D-N.

Table 2. Ten operations became three operations after a scalability triage.

| Operation | Load | Work | Quality threshold |
|---|---|---|---|
| **Payment** | **H** | **H** | ***n* seconds** |
| Operation 2 | H | L | *n* seconds |
| Operation 3 | L | VH | *n* seconds |
| Operation 4 | H | L | *n* seconds |
| Operation 5 | L | M | *n* seconds |
| Operation 6 | L | M | *n* seconds |
| Operation 7 | L | H | *n* seconds |
| **Balance** | **H** | **H** | ***n* seconds** |
| **Transactions** | **H** | **H** | ***n* seconds** |
| Operation 10 | L | ?? | *n* seconds |

## 6.2 Scalability Scenarios

The three different scenarios in our open banking case study were:

- Realistic: Reflecting the current customer base in October 2018, plus an estimate of TietoEVRY's usage of open banking two years later, in October 2020. The focus is on moderate use.
- Possible: Assuming more banks outside the current customer base, plus one of the three large players as TPP for TietoEVRY's customers. The three large players were perceived to be a large national player plus Google and Facebook. In addition, there would be more extensive use of open banking.
- Extreme: The same as the possible scenario, but two of the three large players as TPPs. Even more extensive, but still likely, use of open banking.

In addition to the number of TPPs, the maturity of the market and user behavior are important factors determining the load. The structure and formulas of the model are the same as those used in TietoEVRY, but parameter values are obfuscated. An underlying assumption is that open banking becomes a success and not a flop. Usually, the load will increase slowly, but large players like Facebook may get consent from many users, quickly causing an abrupt load increase.

## 6.3 Generic Input Parameters

Several operations share a set of generic input parameters, shown in Table 3. How reliably these parameters may be estimated varies, but through discussions with key stakeholders, we arrived at reasonable values.

Table 3. Generic input parameters.

| | Description | Realistic | Possible | Extreme |
|---|---|---|---|---|
| $n_t$ | #TPP apps per bank customer | 1.0 | 2.0 | 3.0 |
| $f_a$ | Fraction of active bank customers | 0.3 | 0.5 | 0.8 |
| $c$ | # bank customers granting access to TPPs | 1 000 000 | 2 000 000 | 3 000 000 |
| $a$ | Average # accounts per bank customer | 2.0 | 3.0 | 4.0 |
| $p_m$ | Additional average # payments per bank customer per month | 1.0 | 2.0 | 3.0 |
| $d_m$ | Days per month | 30 | 30 | 30 |

## 6.4   Derived Parameters

The second group of parameters, shown in Table 4, contains derived parameters based on the other input parameters described above.

- Average number of active accounts: $c_a = c * a * f_a$
- Additional average payment operations per hour: $p_h = c * f_a * p_m / (d_m * 24)$

Table 4. Derived input parameters.

|      | Description                         | Realistic | Possible  | Extreme   |
|------|-------------------------------------|-----------|-----------|-----------|
| $c_a$ | # active accounts                  | 600 000   | 3 000 000 | 9 600 000 |
| $p_h$ | Additional average # payments per hour | 417    | 2 778     | 10 000    |

## 6.5   Payment Input and Output Parameters

To derive the load on the payment operation, the concept of burstiness was instrumental. Burstiness is the peak amount of load on some time scale relative to the average load on the same time scale. For example, monthly burstiness, $b_{m,p}$, is the load in the busiest month of the year as compared to the average monthly load, for example, the load in the busy Christmas month of December with an increase in shopping activity and payment transactions, compared to the average month. To find the total hourly burstiness of the year, we simply multiply all the burstiness measures per month, day, and hour.

The load on the payment transaction in terms of the additional maximum number of payments per second, in the busy hour, can now be computed (see Table 5):

$$p_s = b_{m,p} * b_{d,p} * b_{h,p} * p_h / 3\ 600$$

Table 5. Additional payment load.

|        | Description                            | Realistic | Possible | Extreme |
|--------|----------------------------------------|-----------|----------|---------|
| $b_{m,p}$ | Monthly burstiness (bank customer)  | 1.5       | 1.5      | 1.5     |
| $b_{d,p}$ | Fraction of active bank customers   | 2.0       | 2.0      | 2.0     |
| $b_{h,p}$ | # bank customers granting access to TPPs | 2.0  | 2.0      | 2.0     |
| $p_s$  | Additional load per busy second        | 0.7       | 4.6      | 16.7    |

Based on the example parameter values and an expert evaluation, the three scenarios will be relatively easy to handle. The load on these three scenarios is, therefore, marked with green. With open banking, there will likely not be a large number of additional payments. Mainly, payments will move from other platforms, but the total amount will not increase significantly.

## 6.6   Balance Input and Output Parameters

According to the PSD2 directive, the maximum number of balance requests per day per bank customer per TPP has an upper bound, currently four. We also assume that load in the busiest hour is three times the average daily load. An important reason for this number is that TPPs will try to get a fresh balance, and, at least currently, the balance is updated only a few times per day. TPPs will then try to access the balance just after a new balance is available. With more instant payments, the balance may be updated more frequently. As a result, the balance requests may also be more evenly distributed. This may increase the maximum number of TPP requests per day. We estimate a fraction of TPP requests actually taking place. Based on these numbers and hourly TPP burstiness, we can derive the load from TPPs on balance operations per second, $e_{t,s}$:

$$e_{t,s} = n_t * c_a * b_{h,b} * a_d * f_d / (24 * 3600)$$

The similar load from bank customers, $e_{c,s}$, is:

$$e_{c,s} = c_a * a_c / 3600$$

Total balance load $e_s$ is the sum of $e_{t,s}$ and $e_{c,s}$; see Table 6.

Table 6. Balance request load.

| | Balance | Realistic | Possible | Extreme |
|---|---|---|---|---|
| $b_{h,b}$ | Hourly burstiness (TPPs) | 3.0 | 3.0 | 3.0 |
| $a_d$ | Max balance requests per TPP per account per day | 4.0 | 4.0 | 4.0 |
| $f_d$ | Average realistic fraction of TPP requests per day | 0.5 | 0.5 | 0.8 |
| $e_{t,s}$ | Load from installed TPP apps per second | 42 | 417 | 3 000 |
| $a_c$ | Max # balance requests per bank customer per hour | 0.2 | 0.2 | 0.2 |
| $e_{c,s}$ | Load from bank customers per second | 33 | 167 | 533 |
| $e_s$ | Total load per busy second | 75 | 583 | 3 533 |

With the example data we have used in this article, the analysis of the balance operation shows that the realistic scenario is well within the current known performance (green). The possible scenario is probably also manageable but should be tested (yellow), whereas the extreme scenario might be challenging for the current system configuration (red). This color scoring was the result of an expert evaluation.

## 6.7 Transaction History Output Parameter

Derivation and analysis of the load for the operation that fetches transaction information are similar to the load on the balance operation and will not be further described here.

## 6.8 Essence of the ScrumScale Model

When we abstract the concepts used in this instantiation of the ScrumScale model, we obtain the following elements:

- Common input parameters for all operations. These common types of input parameters can be broken down further to:
  - Average numbers, like the average number of TPP apps per bank customer, the average number of accounts per bank customer, and the additional average number of payments per bank customer per month (a small number in the order of 0.1 to 10) and bank customers granting access to TPPs (a large number in the order of millions).
  - Constants, like days per month.
  - Fractions, like a fraction of active bank customers.
  - Burstiness, per hour (for all hours in a day), day (for all days in a month), and month (for all months in a year)
  - The degree of uncertainty in the input parameters may also be helpful, but not in our case.
- Input parameters for each operation
- Derived parameters, based on input parameters
- Output parameters per operation, derived from common input parameters or (derived) input parameters. These output parameters should be intuitive for practitioners.

- Scenario: A collected set of input parameter values and output parameter values that together reflect a business scenario. A short description of the rationale behind each scenario should be written.
- Risk: The value of the output parameters can be colored, reflecting how hard they are to fulfill. In our case, we have scored them three values with green, yellow, and red. More than three colors could be used, of course. We could also have used words like very low, low, moderate, high, and very high (Brataas et al., 2021). This scoring is the result of a subjective expert evaluation. The scalability experts compare the projected load with their knowledge of the capacity for the operations in question.

This ScrumScale model will also apply to our other case studies, as described in Section 9.5. Although this list has been distilled from our experience with this case study, some other contexts might require additional or different parameters according to their specific constraints.

# 7  Findings from Testing

In this section, we describe how performance and scalability testing was performed for open banking. The focus was on performance testing in the production environment. Proper scalability testing in a test environment would have been highly desirable, but time did not allow for this.

The open banking application was tested in a production environment, where security mechanisms are strictly enforced, making it hard to create new test users. As a result, only a very limited number of users can be used, in the order of 15. Underlying systems for actual banks were used during this production testing. During these tests, TietoEVRY found long response times in the underlying systems. The open banking system in itself performed adequately. These long response times, together with the limited number of users, made it hard to increase the load. Finding these problems in the underlying bank systems was valuable for TietoEVRY's bank customers. Later, the latency problems in the underlying bank systems were fixed, but high load could not be tested in the production environment because it might disturb the production workload. As a result of a limited number of test users and a limit in the load from each of these users, it was not possible to do proper scalability testing in the production environment.

We also discussed how to do more extensive scalability testing in a sandbox test environment, where it is easier to generate more test users so that we could test with a more extensive workload. TietoEVRY had challenges with this sandbox test environment because of high response times and poor correspondence between the response times that TietoEVRY saw in their test tools and the response times seen in the logs from open banking. Time did not allow for further sandbox testing.

In summary, TietoEVRY was able to test the performance of the open banking system with a low load. During these tests, too-long response times in the underlying systems were found and fixed. Generally, scalability testing of all the different scenarios may not be possible. Extreme scenarios may simply be so tough that it becomes infeasible to generate the required workload. Enough test hardware may not be available. However, if TietoEVRY were able to test the realistic and the possible scenario, it could build an argumentation explaining why it was feasible to cope with the extreme load. Ideas for further work are sketched in (Brataas et al., 2018a).

# 8  Findings from Development

We used the ScrumScale method (Brataas et al., 2020) to involve the key stakeholders, and we used the ScrumScale model (based on the conceptual ScrumScale framework) to estimate the input and output parameters. The ScrumScale method is formulated primarily for greenfield projects with user stories. Still, it is also applicable to brownfield projects like in this case study, with existing underlying subsystems and existing operations, but where further development is a challenge. Here, we reflect on findings during this development process of the ScrumScale model used throughout the assessment.

First, in Section 8.1 we describe the initial structure of the ScrumScale model, how it was created, and the rationale for the information needed. Then, in Section 8.2 we describe how the ScrumScale model evolved during the phases of the project and what we have learned from the iterative changes. Section 8.3 elaborates on how the ScrumScale model was applied in TietoEVRY.

## 8.1 The ScrumScale Model: Focusing and Prioritization

The information required for the structure of the ScrumScale model was collected by employing 11.5 of TietoEVRY's person/hours from May 2018 until July 2018. Importantly, this structure remained fixed throughout the process. We started with the ScrumScale model in Section, 6 and we wanted to know the workload on the critical operations. Then we iteratively—through discussions with stakeholders—found what was available as input parameters. The questions used to derive this basic structure came from steps B and C in the ScrumScale method in Figure 2 (and in Brataas et al., 2020), briefly described in Section 2. The structure included the following vital information:

1.  **Overall business requirements:** Some of the information presented in the Environment section was collected initially, but most came gradually during the subsequent meetings and iterations.
2.  **Type of users:** We focused on one type of user, namely, typical internet bank users. These users are the main users of open banking. Other internal bank staff users were not relevant for open banking, which is all about opening up to the public and not about internal bookkeeping. An important exception was, of course, the TPPs. In contrast to ordinary internet bank customers, the load from TPPs was much harder to estimate, simply because this was a new type of user.
3.  **Type of workload:** As a consequence of focusing on typical bank users in the previous point, we were interested in online transactions. Batch transaction for internal bookkeeping was not studied.
4.  **System boundary:** For the system boundary, it was important to get access to an architecture sketch showing the internal components of the open banking system and its surrounding components. At this early stage, it was unclear whether or not the underlying internal systems were part of open banking, i.e., if they were components inside or outside the open banking system. This became clear after more discussions with different stakeholders during later iterations.
5.  **Three critical operations:** For this brownfield project, it was essential to access documentation showing all relevant open banking operations. It was also critical with the scalability triage step (step B) in the ScrumScale method to reduce this set of 10 operations to a more manageable set of tree operations.
6.  **The output load parameter** for these three critical operations. From the outset, it was clear that load came from two sources: TPPs and own use. This affected only two of the three critical operations, as TPPs do not do payment transactions.
7.  **Input load parameters** required to derive the expected load on the three critical operations for all three scenarios. Most of these parameters remained in the final version of the ScrumScale model, but some parameters were refined because of improved insight. Initial values for these parameters were also set but were later refined when we involved a broader set of stakeholders.
8.  **Work:** We focused on transaction history and number of transactions retrieved per operation, as this seemed to be the most crucial work object in the heaviest critical operation.
9.  **Quality metric:** Response times are most important, and we decided on 90 percentile response times, as this is the metric most commonly used for this type of application in TietoEVRY.
10. **Quality thresholds:** Because performance should be the same for TPPs as for TietoEVRY's own customers, the response time requirement was the result of TietoEVRY's SLA.

11. **Three scenarios** reflecting several potential sets of input parameters for the output parameter load. We started without scenarios but soon discovered several potential sets of input parameters corresponding to different market forecasts: realistic, possible, and extreme. Some refinement of input data for all these scenarios was done, but the overall structure remained. The names of these three scenarios were also refined.

12. **Consistency:** The quality of the data presented to the TPPs should be the same as the data presented to their own customers. Therefore, simple replication of stale (not-so-fresh) data was not an option. TPPs had to use the same systems as their own customers.

The organization made decisions according to insights that were elicited during the artifact design. For example, the operations to be analyzed and then tested to avoid scalability debt were prioritized. This avoided a heavy upfront approach that would involve the test of all operations in all the subsystems. Some operations were not critical in terms of scalability (for example, having high work but low load). Consequently, subsystems that did not have critical operations did not need to be tested, which would lighten the process of assessing possible scenarios. This information was essential in focusing and prioritizing. It cannot be overestimated how vital a clear focus is, as otherwise, a scalability analysis becomes wholly overwhelming and not feasible, resulting in the fix-it-later approach.

## 8.2   Revising the ScrumScale Model in Later Cycles/Iterations

After the first implementation of the ScrumScale model, it was essential to receive feedback and acceptance from the stakeholders and the management to continue developing and using the ScrumScale model in the organization. We continuously quality-checked the values to plan with the stakeholders how to perform the testing and to be reactive in the face of uncertainty. In fact, given the large number of stakeholders and needs on different levels involving open banking, requirements changes were frequent and had to be considered.

Driven by questions for step 1 in Figure 2 and in the IEEE SW article (Brataas et al., 2020), the set of parameters in the ScrumScale model was refined in sprints D-N. It was helpful to develop the ScrumScale model in an agile way, especially iterating on the information needed along the process. The values of the parameters were scrutinized and updated by different stakeholders. Reflections on how we used the questions in step 1 are as follows:

1. We revisited the system boundary and decided to include the underlying systems when testing, in particular both the Core and the Payment subsystem. However, these underlying systems were not part of the open banking development project. We discussed which of the subsystems were relevant for all banks. Some were relevant only for some banks.

2. For load, several issues were relevant:
   a) We started with transactions (load) per day. When asking for feedback from stakeholders, we found that transactions per second was a more intuitive unit.
   b) Burstiness proved to be a helpful concept, and we gradually experimented with splitting up burstiness in months, weeks, and days. We also tried with active hours per day but realized that burstiness was more intuitive. These two measures are related.
   c) We discussed what it meant to be an active customer and ended up with active on a monthly and not daily basis.
   d) Typically, the load on the critical operations increases slowly, but major players like Facebook may get consent smoothly and move fast. This is important because you must do scalability testing well in advance so that you know your capacity.
   e) The load on the payment operation would not increase so much; instead, the workload would move from other sources.
   f) We experimented with uncertainty in the input parameters so that some input parameters had high uncertainty and others had low uncertainty. This was not very helpful, so we skipped it.

3. Concerning work objects in the transaction history operation, we could not estimate the required input and output parameters. However, we captured the vital insight that the transaction history operation is heavier than the balance operations.

4. Response time requirements did not change. This is natural since response time requirements are derived from a standard service level agreement (SLA).

5. We revisited the three critical operations initially selected in the scalability triage several times during step 1 in different sprints, especially when we introduced the model to new stakeholders, but they did not change.

6. Concerning communication with the product owner, we discussed if, during high load, we should use throttling/rejection to take care of our own customers. This would relax response times or the freshness of data (using caching, for example). This may be a violation of the PSD2 directive, but in cases of emergency, it is crucial to have a clear view beforehand, of what should be relaxed. We also discussed replication options for the Core DBMS.

7. The scenarios were refined. We started with a *realistic* scenario and two *possible* scenarios. Later, it was found to be more beneficial to exclude the realistic scenario, as, on closer examination, we found that it did not tell us about the future but, instead, about the present. We added an *extreme* scenario and renamed the simplest possible scenario into a realistic scenario. We, therefore, ended up with a realistic scenario, a possible scenario, and an extreme scenario.

8. Some information, which was deemed helpful in the beginning, was found to be superfluous. In particular, it turned out that the number of TPPs did not affect the load on any of the critical operations. This was surprising, but it turned out that the average number of TPP apps for each potential customer was the essential parameter in this respect. On the other hand, other input parameters became more important—for example, the average number of TPP apps per customer.

The analysis of the scalability requirements for the testing triggered the investigation of possible architectural solutions to avoid architectural debt. We found it interesting, as the artifact allowed for identifying potential solutions at an early stage.

## 8.3    Application of the Model

This section describes how the ScrumScale model and spreadsheet were applied in TietoEVRY. The model was both a practical tool and a compact representation of reality. The model structured the discussion of the scalability requirements in several meetings. When we changed inputs to the model, it was easy to see the effects on the outputs. It also became clear which input values were important. TietoEVRY did not have this type of discussion in such a structured way before. In these meetings, the model was applied as well as developed.

Early sharing of the model made the stakeholders aware of possible implications for development and testing. Initially, TietoEVRY planned to exclude subsystems like Core and Payment during performance testing of the open banking project. Based on discussions around using the spreadsheet, they decided to include these systems in the testing. As a result of this production testing, long response times for one particular bank were identified, deep down in the customer's solution's subsystems, as previously described in Section 7.

The model was also used to prioritize testing. The balance and transaction history operations with scalability challenges used the Core subsystem. As a result of the high load on the Core subsystem, the persons responsible for this subsystem decided to do more extensive performance testing. We describe the Core subsystem in more detail in Section 9.2. In Section 9.2, we further explain how the triage process was helpful in the Payment subsystem development.

# 9    Findings from Evaluation

Capturing scalability requirements is a complex task. This section will describe the value of the ScrumScale method, the ScrumScale framework, and the ScrumScale model. We will describe the relation to operations and what is required to benefit from the ScrumScale framework and the ScrumScale model. Finally, we will also explain the benefits and challenges in other case studies.

Our evaluation was conducted both with persons who had assisted in developing the ScrumScale model and those who were only marginally involved. This assured that we received feedback not only from the ones who had invested in the artifact but also from less invested, and therefore biased, stakeholders. For example, the head of non-functional testing was only present in the kick-off for this case study. He had earlier been mildly skeptical towards the benefit of ScrumScale in TietoEVRY. The project manager for the open banking project was only marginally involved in creating of the ScrumScale model. The chief technical architect and team responsible for the Core subsystem (the same person) was only marginally involved in creating the ScrumScale model. This was also the case for the chief technical architect for the Payment subsystem.

We observed a change in attitude towards creating a lightweight artifact; initially, some of the core stakeholders expressed skepticism that changed over time, where they increasingly got more engaged in the work. In addition, the number of stakeholders involved in using the artifact grew over time with a "snowballing effect," which could only be possible if they received value by using the model. We interpret these facts as a confirmation of the usefulness of the artifact for its stakeholders and users.

## 9.1    Value of the Artifacts

**Value of the ScrumScale Method:** It is challenging to estimate load profiles for new products with no operations statistics. The head of non-functional testing states: "*One of the largest problems for performance (and scalability) testing is the lack of early involvement. Often, we are not consulted until the architecture is completed, all use cases are written, and functional testing is almost completed.*" This case study was one of the few examples of early involvement. He continues: "*Examples of inferior scalability are too little known internally.*" Better knowledge of the cost involved in inferior scalability will increase the willingness to pay for it.

TietoEVRY has benefited from scalability triage. It is not necessarily the heaviest or most frequent transaction that creates problems, but the transactions in which the product of work and load is the highest. Using scalability triage, scalability testing may focus on high risks in the Core subsystem. "*Even more importantly, this will give more dedicated personnel, as they find more meaning in the job they do,*" according to the head of non-functional testing. In Section 9.2 we describe how the triage process was helpful in the Payment subsystem development.

**Value of the ScrumScale Framework:** According to the project manager, the concepts in the ScrumScale framework have increased the awareness of scalability in the TietoEVRY organization: The head of non-functional testing states: "*The concepts in ScrumScale have given big advantages in dialogues about scalability testing with other stakeholders.*" In particular, the concepts of work, load, and workload, when linked to scalability, have given insights to the company practitioners, but even more importantly, when communicating with stakeholders that are more distant to scalability testing.

**Value of the ScrumScale Model:** The software architect compared the ScrumScale model to "*a scalability version of an API definition, a WSDL file, or a Swagger file.*" With the ScrumScale model, it is easier to develop meaningful scalability objectives for scalability testing, which reflect the anticipated open banking workload. As a result, "*we have never had this focus (on scalability objectives) so early before*," according to the main scalability tester. Thus, the underlying subsystems were prepared earlier. This work increased the awareness of proper scalability testing, as well as of the subsystems, so that they discovered high response times deep down in the subsystems of the customer's solution, as already mentioned in Section 8.3.

We asked the main scalability tester the following question: How do we know that these estimates are good enough? He replied: *"We can't, obviously, but they are in my opinion founded on sound reasoning, not overly complicated calculations, and with estimating both realistic, possible and extreme scenarios, we've covered a wide array of cases, and the model also lets us tweak the various input parameters and see what effect that has on the total traffic numbers."*

The project manager said: *"I have worked here for a long time, but we have never done this in a systematic way (until now)."* The ScrumScale model gave *"high load values with a good justification."* This helped to put a focus on scalability requirement elicitation and subsequent scalability testing. As a result of the ScrumScale model, it is also possible to work with scalability requirement elicitation in smaller chunks and not only once for a large system. The project manager stated, *"This could be the way to introduce scalability testing more properly when initiating a project."*

**Relation to Operations:** TietoEVRY has an active capacity management policy with daily supervision and weekly capacity management meetings. The numbers in the ScrumScale model will be valuable for the persons in operations because they now *"have a better feeling for which workload it would be possible to expect."* This has consequences for the metrics they monitor and makes it easier for operations to do a good job. In this way, scalability testers contribute to operations.

TietoEVRY has not yet fully implemented DevOps, which means the scalability tester is seldom part of the team. The ScrumScale model may be the communication vehicle required for performing more DevOps. It is often argued that *"since our operations environments are dynamically scalable, we can add CPUs or more servers if this is required."* The underlying assumption is that the system can exploit these resources, i.e., that the system is scalable. The head of non-functional testing states, *"We have general requirements that the solution should scale horizontally and vertically, but without a method for actually testing it."*

## 9.2   Further Benefits of ScrumScale in TietoEVRY

This section describes how TietoEVRY has used the ScrumScale method in two of the most vital underlying open banking subsystems, first for Payment and afterward for Core functionality. Toward the end of this section, we outline the further potential of the ScrumScale method in TietoEVRY. This, of course, includes the ScrumScale framework and the ScrumScale model.

The Payment subsystem has used the triage technique with success. With this risk evaluation technique, it was possible to speed up the increment frequency to one release per month, as performance testing was required only for the features with high risks. This prioritization also allowed for more performance testing of these risky parts of the solution.

One of the goals of the large Core project is to leverage new technology to achieve, among other things, better performance and scalability. As is so often the case, the primary focus from the start was on the functionality of the system, relegating performance issues to the conventional fix-it-later approach (Smith and Williams, 2001), as described in the state-of-the-art Section 2. When performance became an issue later in the project cycle, costly refactoring efforts had to be initiated.

The performance evaluation carried out in the Core project focused on response times under normal load. In fact, the load is not part of the SLA (service-level agreement) with the customers, so, implicitly, the response times should be the same regardless of load. It should be possible to scale the system, but this scaling is not specified, except for the general phrase "the system should be able to scale both horizontally and vertically." With no specification of scalability, it will also be hard to test scalability.

As a basis for performance testing in the Core project, a spreadsheet like the Open Banking scaling model was used but focusing only on one realistic scenario. Production data from the existing system was used for load. For work, the complexity of transactions was used. Business criticality was used instead of the quality threshold. The triage approach could have been used here. In retrospect, the project manager concluded that targeting scalability, not only performance, and exploring a set of scenarios, not only the most realistic one, would have benefited the project.

The solution architect of the Core project confirmed that an incremental approach, as suggested by ScrumScale, would have been beneficial to steer the project in the right direction earlier. Recently, the project has adopted a more incremental way of working, and it is easy to see that there is now a better match with the ScrumScale method. The chief architect and team responsible for the Core project (the same person) concluded the interview as follows: "*I have a great sense of what you have done, and I think it is excellent! But it has to be planned, and it has to be worked on to make it fit.*"

## 9.3   Benefits and Challenges in Other Case Studies

In the context of ScrumScale, we also did several other case studies: 1) credit card accounting in TietoEVRY, 2) intraday energy trading in Energy, and 3) management of building applications in Altinn. Here we will briefly describe the benefits and challenges of the latter case study, the Building Application Case Study, which illustrates the ScrumScale method (Brataas et al., 2020). In the Building Application Case Study (BACS), we investigated the scalability requirements for a building application system for the most prominent public portal in Norway, Altinn. Whereas the load parameters represented the most challenging scalability requirements in the open banking case study, the work parameters were more important in BACS. In BACS, clarifying the scalability requirements led to several important insights: (1) The development team could prepare Altinn in time for the larger-than-expected building applications. (2) A service hotel could be designed to handle some of the unanticipated but related workload from neighbor notifications. (3) Moreover, a requirement for changing applications in the middle of the application process was relaxed, as it would lead to an extreme workload.

Ten different organizations were involved in setting functional requirements, making different parts of the service, modifying the platform, managing development, performing operations, etc. The amount of information was overwhelming because several of these stakeholders had been involved for years. Using the lightweight ScrumScale method, we contributed valuable advice with a minimum of involvement. In total, BACS stakeholders spent 50 person hours in meetings and handling emails with follow-up questions. Training sessions on the ScrumScale method were an integral part of these meetings.

# 10   Discussion

We first report how we have answered our research questions. Then we outline our contributions and lessons learned in the fundamental research fields. Finally, we discuss the limitations of the study and how we mitigated existing threats to validity.

## 10.1 Answers to Research Questions

In this study, we report how a lightweight artifact and method was developed and evaluated in a real-world case to solve the challenges posed by a business environment (in the financial domain) with evolving scalability requirements and uncertainty (open banking).

**RQ1: Is it feasible to use an iterative and lightweight approach to create an effective artifact to elicit scalability requirements?**

The lightweight ScrumScale model was successfully developed with limited resources as we consumed only 55 hours of TietoEVRY stakeholder time. The model was also found valuable by the organization (see RQ2), showing that the artifact was effective. This provides evidence that, in the specific organization, we can positively answer RQ1.

Using a design research process approach, we also report the process and the lessons learned during the development and application of such artifact, which shows *how* it was made feasible. For example, we report the various refinements necessary to develop and apply the artifact, the involved roles, and the effort needed for its development, the challenges, and the enablers. One important outcome is the validation and further knowledge on using the first sprints of the ScrumScale method to support the usage of the ScrumScale model.

The lessons learned reported here can be used and applied by similar organizations facing similar challenges (openness and uncertainty) concerning scalability requirements to create a similar artifact, as mentioned in contribution C4 in Section 1. By carefully describing the environment and how the method was developed, we believe that such an approach can be tailored to other contexts. To assist this task, we list other cases where the approach was applied, and we report corresponding benefits and challenges.

We speculate that the method could be tailored to cover the elicitation and analysis of other NFRs. However, we believe that such a step would undoubtedly require several adaptations of the artifact and, therefore, further studies would be needed. However, our work can be used as a first step to guide such work.

**RQ2: How well did our artifact support agile scalability requirement elicitation for open systems?**

After developing and applying the ScrumScale model, we report the benefits and challenges of using the artifact. We evaluated the model in a real-world case to solve the challenges posed by a business environment (in the financial domain) with evolving scalability requirements and uncertainty (open banking). Such a model helped successfully coordinate the stakeholders' input in managing scalability, improving the organization's current practice, as shown by our results from the qualitative interviews. In particular, the ScrumScale model helps with eliciting and continuously assessing scalability requirements according to the most relevant parameters or, in other words, constitutes a viable *lightweight coordination* mechanism (spanning object).

The ScrumScale method, the ScrumScale framework and the ScrumScale model was valuable for the organization, as it helped improving its practices by:

1. Providing a systematic approach that could be developed and applied in a lightweight way, i.e., using limited resources.
2. Improving collaboration across various scalability stakeholders (project managers, architects, head of non-functional testing, testers, etc.) using a lightweight coordination mechanism, the ScrumScale model.
3. Contributed to learning about open banking, e.g., that the number of third-party providers (TPPs) did not affect the open banking workload, as stated in Section 8.2. The three scenarios, realistic, possible, and extreme, clarified the thinking. Major players like Facebook may get quick consent from many users, causing abrupt open banking workload increases.
4. As a concrete output from this case study, Section 6.8 provides the essence of the ScrumScale model. This essence will hopefully make it easier for others employing a similar approach.
5. Increasing awareness in the organization of the scalability concerns using the ScrumScale framework. As described in Section 9.1, this gave advantages in dialogs on scalability testing with other stakeholders.
6. Assisting the prioritization of relevant systems to be tested, hence saving precious resources using the ScrumScale model. It helped them know that the projected workload on the Core subsystem could be high. The Core subsystem served the balance and transaction history operation. The Payment subsystem's workload, used by the Payment operation, would be reasonably stable with the introduction of open banking. As a result, more performance and scalability testing were relevant for the Core subsystem.

7. An earlier and precise focus on scalability testing makes the testing job more exciting and increases work motivation, as described in Section 9.1. Using the ScrumScale model, it was possible to develop meaningful scalability objectives for scalability testing.

8. Operations were more prepared for which workload to expect. This work may also increase the cooperation between operations and testing.

9. Outside of this case study, the scalability triage step (step B) in the ScrumScale method helped speed up the increment frequency in the Payment subsystem to one release per month, as described in Section 9.2. This illustrates how the TietoEVRY organization has already adopted the ScrumScale method.

## 10.2 Other Contributions

Besides answering our research questions, our study is at the intersection of various research fields. This is often the case when one designs an artifact that must solve a practical problem. Therefore, we also find the following contributions for various areas of software engineering research.

- Agile Scalability Requirements Engineering

Our research contributes to agile scalability engineering. We describe the ScrumScale model and illustrate how it was used to manage scalability requirements engineering in TietoEVRY. We show how the ScrumScale model was used as a lightweight spanning object assisting communication in TietoEVRY. This communication was performed during short meetings with key stakeholders present. All stakeholders did not participate in all meetings, but the ScrumScale model was gradually instantiated during a series of short meetings. We also describe the ScrumScale framework in more depth so that it and the ScrumScale model can be used in other contexts. The initial steps in the ScrumScale method were validated (steps B, C, and 1). We showed the power of the scalability triage step (step B) to focus the effort on the features with significant scalability risks.

- Coordination in agile architecture

Our study contributes to agile architecting, especially to the coordination aspect, with a concrete method and tool used in practice to assess scalability in an iterative and lightweight way and to identify risks in the presence of uncertainty.

The ScrumScale model has been proven effective in fostering agile communication and coordination between architects, testers, and other stakeholders. In particular, we provide a concrete example of how to create and use a spanning activity (the ScrumScale method), a spanning object (the ScrumScale model), and a spanning role (performed by scalability experts) in a large agile project.

The usage of the model has also helped elicit tacit knowledge from the stakeholders. Tacit knowledge is the kind of knowledge that is acquired through experience. Codifying the tacit knowledge from the scalability experts and the stakeholders into the model was key to eliciting scalability requirements.

- Architectural debt avoidance

Our results can also be helpful in avoiding dangerous technical debt in the form of scalability debt, as reported in our previous exploratory work (Hanssen et al., 2019). Architecture technical debt theory advocates that the debt is worth avoiding if the cost of avoiding it is less than the negative impact avoided (Martini and Bosch, 2016b). In this study, a limited number of hours (55) was invested to avoid a massive impact on scalability requirements not met when the system would be exposed to TPPs. Although we do not have a quantified amount of interest (negative impact) avoided, thanks to using the ScrumScale model, our evaluation at the end of the study shows how the stakeholders recognized the approach as valuable and worth the time spent on its implementation.

## 10.3 Limitations and Threats to Validity

The evaluation of the approach showed that it was successful in TietoEVRY to coordinate scalability management in a lightweight way. Partial scalability testing was also conducted. However, not all the scenarios could be evaluated because of constraints related to resources and testing policies. In addition, the method is sensitive to good experts' estimations, in common with other software development activities that are performed at the beginning of a project, such as architecting and cost estimations. In our study, the scalability experts, as a spanning role, use the ScrumScale model as a spanning object to support their contributions. Thus, although we cannot know for sure if the elicited values were realistic, we have gathered evidence from our evaluation interviews that our approach will help TietoEVRY further perform such testing and manage scalability in its everyday work. In addition, we discuss the main threats to validity taken from Runeson and Høst, 2009.

**Construct validity** - The scalability constructs (work, load, etc.) were not measured but were subjectively estimated by the experts. However, the estimations were based on a theoretical framework, the ScrumScale framework, outlined in Section 3. The process was thoroughly followed by an expert in the field of scalability to ensure that the constructs were well developed. The input was refined continuously, and it is essential to note that our method allowed for broad participation in the scalability elicitation process, which would have mitigated the bias of having only one person estimate the values. In addition, with more practice, the quality of the numbers will increase with the next application of the method.

**Internal validity** - In this study, we did not measure causal relationships. However, we claim the approach to be successful or that applying the method brought some benefits to the organization. Although there might be external factors for which the process was considered erroneously successful, by thoroughly following the development and evaluation of the approach, we can reasonably exclude substantial external effects.

**External validity** - Probably the more substantial threat is that we have thoroughly reported on only one case in which our approach was successful. It is, therefore, natural to wonder if and how such an approach would work in other contexts. Considering our literature review, in which we observed that scalability requirements are not usually handled in a lightweight way, we can consider TietoEVRY as a representative organization where scalability requirements were not initially handled in an agile way but were then successfully managed thanks to the development and application of the ScrumScale model, method, and framework.

To mitigate this threat of our findings, first, we have reported a detailed description of the environment where the approach was used. This can help organizations like TietoEVRY apply a similar approach. Second, we report how the method and artifacts were created so that an organization can replicate the *process* of building such lightweight approaches rather than use our method off the shelf (which could not work). Third, we have also successfully applied the scalability triage step to the Payment systems at TietoEVRY, as described in section 9.2. Forth, we have previously reported experiences (benefits and challenges) from one other organization where the method was applied in parallel to the case described here (Brataas et al., 2020). This case study is also briefly described in Section 9.3. This can help practitioners transfer our lessons learned to their projects and domains.

**Reliability** - Given the qualitative nature of our study, it is imperative to ensure that the results are reported reliably, avoiding biased interpretations. Although altogether avoiding subjectivity (both concerning the subjective evaluation from the interviewees and the interpretation of the researchers) is not possible, the following measures were taken to avoid bias where possible: 1) During data collection, on some occasions, more than one of the authors (in a few interviews, three of them) were present. In addition, the findings were frequently quality-checked by the stakeholders in the company. This can be considered an approach to achieve better source triangulation. 2) During the analysis, two authors went through the available data and double-checked the findings independently.

# 11 Conclusion and Further Work

In this case study, we have shown the application of the ScrumScale method and the ScrumScale framework. These are contributions C3 and C2, respectively, as described in Section 1. We see a value in being close to the actual challenges in agile scalability engineering. A common misconception about agile methods is that documentation is not needed and that requirements only should be managed and matured throughout the process. While it is true that agile methods emphasize a lightweight process being guided by the users' needs, our case demonstrated the value of documenting and analyzing non-functional requirements upfront, in this case, related to scalability—however, as a lightweight approach. The approach is supported by an artifact, the ScrumScale model, which has been valuable in this case study. The ScrumScale model corresponds to contribution C1 in Section 1. We also report in detail on lessons learned using the ScrumScale model, corresponding to contribution C4 in Section 1.

In TietoEVRY, the ScrumScale method is already applied in the Payment subsystem outside of this case study. Apart from this, TietoEVRY moves toward more agile practices and plans to exploit further the ScrumScale method, including the ScrumScale framework and the ScrumScale model. The ScrumScale method advocates iterative testing, and more automated performance testing is perceived as low-hanging fruit, enabling more iterative testing.

Further work may be done in at least the following areas:
- Refine the ScrumScale model:
    - It may be more representative so that it fits more systems. Wohlrab et al., 2014, is a source of inspiration for extending the ScrumScale model.
    - In retrospect, it would have been helpful to record precisely who gave each input parameter and who assessed the risks. This is important for credibility, but also because otherwise, it is too easy to say afterward (for example, when the derived load for a scenario was too high), "I do not trust these numbers." However, we must remember that people always resist guessing numbers, precisely because it is easy to blame persons, trying to guess numbers faced with uncertainty. In the bleak light of afterthought, these numbers could be less misleading, which may subsequently have a considerable economic impact. This will increase the burden of stating numbers.
    - Further investigations to evaluate the feasibility and value of the ScrumScale model more rigorously.
- The ScrumScale framework should be further explored. Are these concepts representative for other systems?
- Refine the ScrumScale method: How reliable are the decisions at the scalability triage step? How to handle the situation during triage when the amount of resources varies between the operations?
- More case studies would be valuable:
    - It would be fascinating when other than the authors work as "spanning roles." How do they succeed? Which information do they need?
    - It would be interesting to triangulate the estimations collected through our method with scalability testing to validate our approach further. Then, scalability testing may also be better described in the ScrumScale method.
    - Covering system's operation with more details.
    - How to handle the situation if the load and the work depend on each other?

- The focus should be extended to other NFRs. In addition to scalability, we extend the triage step from scalability to include security, safety, and availability (Brataas et al., 2021). This work can include other method steps.

## Acknowledgments

The research leading to these results has received funding from the Research Council of Norway; ScrumScale: Scalability in Agile Software Development (grant #256669), SMED: Smarter Innovation with Digital Transformation of Innovative Procurement (grant #285542), and CapGuard: Capacity Guard of Public Digital Services (grant #309811). We thank TietoEVRY for their hospitality, which made this case study possible through access to this business's critical open banking project and providing access to critical stakeholders and vital documentation.